\documentclass[aps,twocolumn,prl,superscriptaddress,longbibliography,reprint]{revtex4-2}
\usepackage{amsfonts}
\usepackage{mathrsfs}
\usepackage{amsmath}
\usepackage{color}
\usepackage{graphicx}
\usepackage{bm}
\usepackage{amssymb}
\usepackage{xspace}
\usepackage{epstopdf}
\usepackage{longtable}
\usepackage{multirow}
\usepackage{booktabs}
\usepackage{textcomp}
\usepackage[title]{appendix}
\usepackage{float}
\usepackage{dsfont}
\usepackage[colorlinks=true, letterpaper=true, pdfstartview=FitV, linkcolor=blue, citecolor=blue, urlcolor=blue]{hyperref}
\usepackage[]{changes}
\usepackage{overpic}
\usepackage{makecell}
\usepackage{adjustbox}

\providecommand{\U}[1]{\protect\rule{.1in}{.1in}}

\begin{document}

\title{Out-of-Plane Nonlinear Orbital Hall Torque}

\author{Hui Wang}
\thanks{These authors contributed equally to this work.}
\affiliation{Science, Mathematics and Technology (SMT), Singapore University of Technology and Design, Singapore 487372, Singapore}

\author{Xukun Feng}
\thanks{These authors contributed equally to this work.}
\affiliation{Interdisciplinary Center for Theoretical Physics and Information Sciences (ICTPIS), Fudan University, Shanghai 200433, China}

\author{Jin Cao}
\affiliation{Research Laboratory for Quantum Materials, Department of Applied Physics, The Hong Kong Polytechnic University,
Kowloon, Hong Kong SAR, China}

\author{Huiying Liu}
\affiliation{School of Physics, Beihang University, Beijing 100191, China}

\author{Weibo Gao}
\affiliation{Division of Physics and Applied Physics, School of Physical and Mathematical Sciences, Nanyang Technological University, Singapore 637371, Singapore}
\affiliation{School of Electrical and Electronic Engineering, Nanyang Technological University, Singapore}
\affiliation{Centre for Quantum Technologies, National University of Singapore, Singapore}

\author{Cong Xiao}
\email{congxiao@fudan.edu.cn}
\affiliation{Interdisciplinary Center for Theoretical Physics and Information Sciences (ICTPIS), Fudan University, Shanghai 200433, China}

\author{Shengyuan A. Yang}
\email{shengyuan.yang@polyu.edu.hk}
\affiliation{Research Laboratory for Quantum Materials, Department of Applied Physics, The Hong Kong Polytechnic University,
Kowloon, Hong Kong SAR, China}

\author{Lay Kee Ang}
\email{ricky\_ang@sutd.edu.sg}
\affiliation{Science, Mathematics and Technology (SMT), Singapore University of Technology and Design, Singapore 487372, Singapore}

\begin{abstract}
Despite recent advances in orbitronics, generating out-of-plane orbital torques essential for field-free deterministic switching of perpendicular magnetization remains a key challenge. Here, we propose a strategy to produce such unconventional torques across broad classes of materials, by leveraging the nonlinear orbital Hall effect.
We demonstrate that this nonlinear orbital response is dramatically amplified by topological band degeneracies, where it overwhelmingly dominates the spin response even in systems with strong spin-orbit coupling.
These features are confirmed via a quantitative investigation of representative topological metals  RhSi, YPtBi, and PbTaSe$_2$, by combining our theory with first-principles calculations. The resulting orbital torques substantially surpass those from linear mechanisms reported thus far. These findings propel the research of orbital transport into the nonlinear regime, broaden the scope of orbital source materials, and establish a new pathway towards high-performance orbitronic devices.

%

%
\end{abstract}
\maketitle

Orbital Hall effect, which generates a transverse flow of orbital angular momentum by an applied electric current, has been a focus of recent research~\cite{bernevig2005orbitronics,Guo2005,kontani2009giant,go2018intrinsic,jo2018gigantic,Bhowal2020,sala2022giant,Salemi2022PRM,Manchon2022OHE,Rappoport2023orbital,choi2023observation,Kawakami2023OHE-Cr,Mertig2024OHE,Mertig2025OHE}. As a potential application, its produced orbital current can be injected into a neighboring magnetic layer and exert a torque on the magnetization. This orbital Hall torque offers an electric means to  induce magnetic dynamics or even achieve magnetic switching~\cite{choi2023observation,lee2021orbital,DingPRL2024,Jiang2024OT-PMA-Zr,Jiang2025orbital-FGT}.
%
%
Recent studies have demonstrated strong orbital Hall effect generated by light elemental transition metals~\cite{choi2023observation,Kawakami2023OHE-Cr}, in which the driving charge current, generated orbital current and its orbital polarization are mutually orthogonal. Unfortunately, this type of orbital current is inefficient for manipulating magnets with perpendicular magnetic anisotropy (PMA), which are a key element for high-density scalable magnetic memory devices~\cite{manchon2019,Ryu2020SOT,miron2011,Liu2012,kurebayashi2017view,Yang2023briefing}.

The torque required to achieve field-free deterministic switching of PMA magnets should be out-of-plane, which arises from an unconventional collinearly-polarized (CP) component of orbital Hall current $j^p_a$, namely, the flow direction $a$ and the orbital polarization direction $p$ should be the same~\cite{Ryu2020SOT,Yang2023briefing}. However, such CP orbital current is forbidden by the high crystalline symmetry in the most studied orbital Hall materials, like Ti and Cr~\cite{sala2022giant,choi2023observation,Kawakami2023OHE-Cr}. This has posed a critical challenge for the development of orbitronics.


In this work, we propose a strategy for generating CP orbital currents across broad classes of materials, by exploiting a new physical effect: the nonlinear orbital Hall effect (NOHE). Here, `nonlinear' means the generated orbital current is quadratic in the driving electric field, $j^p_a\propto E^2$. We show that NOHE is far less constrained for producing CP orbital currents: it is permitted in 19 out of the 21 noncentrosymmetric crystal classes. Notably, in 9 of these classes, the linear CP orbital Hall response is entirely forbidden. These nine classes encompass numerous topological materials of recent interest, such as chiral multifold fermion semimetals and half-Heusler topological semimetals.
We develop the theory for NOHE, with both intrinsic and extrinsic mechanisms. We show that the intrinsic NOHE is significantly amplified by topological band crossing features, where the orbital response dominates over the spin response. These points are confirmed by quantitative evaluation in
concrete topological metals: RhSi, YPtBi, and PbTaSe$_2$, by combining our theory with first-principles calculations. Furthermore, we uncover that (i) the out-of-plane torque efficiency from NOHE can surpass that of previously reported linear mechanisms; (ii) while current orbitronic research mainly focuses on light materials with weak spin-orbit coupling (SOC), NOHE is highly sensitive to band topology and can prevail over spin responses even in strong-SOC systems; and (iii) the sign of the NOHE-induced orbital torque can be well controlled via crystal symmetry. Our findings unveil NOHE as a new physical effect, open a new avenue of research for orbitronics, and provide a pathway to overcome the challenges in generating out-of-plane orbital torques.

{\emph{\textcolor{blue}{CP nonlinear orbital Hall current.}}}
The orbital current $j_a^p$ induced by nonlinear response can be expressed as
\begin{equation}\label{chi}
j_a^p = \sum_{b,c} \chi_{abc}^p E_b E_c.
\end{equation}
The nonlinear orbital conductivity tensor  $\chi_{abc}^p$ is a time-reversal-even rank-4 pseudo-tensor for nonmagnetic materials, and requires broken inversion symmetry.
We are interested in the CP current from NOHE, which corresponds to $\chi$ components with $p=a \neq b, c$.
From symmetry analysis, we find that the CP NOHE response is permitted in 19 out of the 21 noncentrosymmetric crystal classes. Particularly, in 9 of these classes, including $C_{2v}$, $D_2$, $D_4$, $D_6$, $D_{2d}$, $D_{3h}$, $T_d$, $T$, and $O$, NOHE provides the leading-order contribution, since the linear CP orbital current
is symmetry forbidden (details in Supplementary Note 1). We shall choose three representative material examples for concrete studies, which belong to $T$ (RhSi),  $T_d$ (YPtBi), and $D_{3h}$ (PbTaSe$_2$). The symmetry allowed components $\chi_{a(bc)}^{a}$ ($\equiv (\chi_{abc}^{a}+\chi_{acb}^{a})/2$) for these 3 classes are listed in Table~\ref{materials_symmetry}.

\emph{\textcolor{blue}{Comparison with spin current response.}}
The orbital current response is always accompanied by a spin counterpart, since the two share the same symmetry.
Before detailed evaluation, one can actually get some qualitative understanding, especially on their relative strength.

The orbital and spin current operators can be expressed as \cite{sinova2015} $\hat{j}_{a}^{p}=\frac{1}{2}\{ \hat{v}_{a},\hat{\mathcal{L}}^{p}\}$ and $\frac{1}{2}\left\{ \hat{v}_{a},\hat{s}^{p}\right\}$, respectively.
Here, the matrix element of orbital angular momentum $\mathcal{L}$ in the band representation reads (take $e=\hbar=1$) \cite{Souza2023multipole,Manchon2022OHE,Manchon2023OHE,Mertig2023OHE,Mertig2024OHE,Mertig2025OHE}
\begin{equation}
\boldsymbol{\mathcal{L}}_{mn}(\bm k)=\frac{i}{4\mu
_{B}}\sum_{\ell\neq m,n}\Big(\frac{1}{\varepsilon_{\ell}-\varepsilon_{m}}+\frac{1}{\varepsilon_{\ell}-\varepsilon_{n}}\Big)\boldsymbol{v}_{m\ell}\times\boldsymbol{v}_{\ell n}, \label{L}%
\end{equation}
where $\mu_{B}$ is the Bohr magneton,
$\varepsilon_n$ is the band energy with $n$ the band index, $\boldsymbol{v}_{\ell n}$ is the velocity matrix element, and the $k$-dependence of the band quantities is made implicit.

One can make the following observations. First, due to the band energy difference factors in the denominator, $\mathcal L$ is significantly enhanced at band near-degeneracies. This suggests that topological metals, which have protected band degeneracies around Fermi level, could be good platforms to realize large NOHE. Second, nonzero $\mathcal L$ as well as orbital Hall response does not require SOC. However, in nonmagnetic systems, SOC is essential for the spin current response, as nonzero matrix element of $s$ requires spin-polarized bands. Third, the diverging factors in $\mathcal L$ do not appear in $s$, implying that the orbital response may dominate over its spin counterpart around band near-degeneracies, even in the presence of strong SOC. Consider a band near-degeneracy with local gap $\Delta\varepsilon$, and $s\sim 1$ for fully spin polarized band. The orbital-spin ratio $\mathcal{L}/s \sim m_e v_F^2/\Delta\varepsilon$, with $v_F$ the characteristic velocity of the electrons.
For typical value of $v_F\sim 10^6$ m/s and relatively large $\Delta \varepsilon\sim 100$ meV, this ratio is already a few tens. This indicates the possible dominance of orbital response in topological metals, even for those having large SOC.

\emph{\textcolor{blue}{Intrinsic NOHE.}}
The features observed above are general, regardless of specific microscopic mechanisms. Below, we shall focus on the intrinsic mechanism of NOHE, which represents an inherent property of each material, determined solely by the band structure.

Employing the extended semiclassical theory \cite{gao2014,gaoPRB2015,gao2019,xiao2021adiabatic}, which has been successfully applied in studying various nonlinear response effects, we obtain the following formula of response tensor for intrinsic NOHE (see Appendix):
\begin{equation}\begin{split}\label{key1}
  \chi_{a(bc)}^{p}=&\frac{1}{2}\int [d\bm k] \Big[\Lambda_{abc,n}^{p}f_0
  \\
  &-\Big(\langle j_a^p\rangle_n G_{bc,n}-\langle v_b\rangle_n \mathfrak{G}_{ac,n}^p-\langle v_c \rangle_n \mathfrak{G}_{ab,n}^p\Big)f_0'\Big],
  \end{split}
\end{equation}
where $[d\bm k]$ is a shorthand notation for $\sum_n d\bm k/(2\pi)^3$, $f_0$ is the Fermi distribution, $\langle v_a\rangle_n=\langle u_{n}|\hat{v}_a|u_{n}\rangle$ is the intraband velocity matrix element for band $n$, similar for $\langle j_a^p\rangle_n$. The tensors $G$, $\mathfrak{G}$, and $\Lambda$ are respectively the Berry connection polarizability, orbital BCP, and orbital BCP dipole. Their detailed expressions are given in Appendix. Like Berry curvature, these quantities are band geometric properties, manifesting interband coherence.


{\renewcommand{\arraystretch}{1.6}
\begin{table}[t]
\caption{Symmetry-allowed nonlinear tensor components $\chi_{a(bc)}^a$ for the representative materials studied in this work. For the point group (PG) $D_{3h}$, the two-fold axis is taken along $y$.}
\label{materials_symmetry}
\begin{adjustbox}{width=\linewidth}
\begin{tabular}{l c c}
\hline \hline
Material & PG & Allowed $\chi_{a(bc)}^a$ \\
\hline
RhSi & $T$ & $\chi_{xyy}^x = \chi_{yzz}^y = \chi_{zxx}^z$, $\chi_{xzz}^x = \chi_{yxx}^y = \chi_{zyy}^z$ \\
YPtBi & $T_d$ & \makecell[c]{$\chi_{xyy}^x = \chi_{yzz}^y = \chi_{zxx}^z$\\ $= -\chi_{xzz}^x = -\chi_{yxx}^y = -\chi_{zyy}^z$} \\
PbTaSe$_2$ & $D_{3h}$ & $\chi_{y(xz)}^y$ \\
\hline \hline
\end{tabular}
\end{adjustbox}
\end{table}}

\begin{figure}[t!]
\centering
\includegraphics[width=0.50\textwidth]{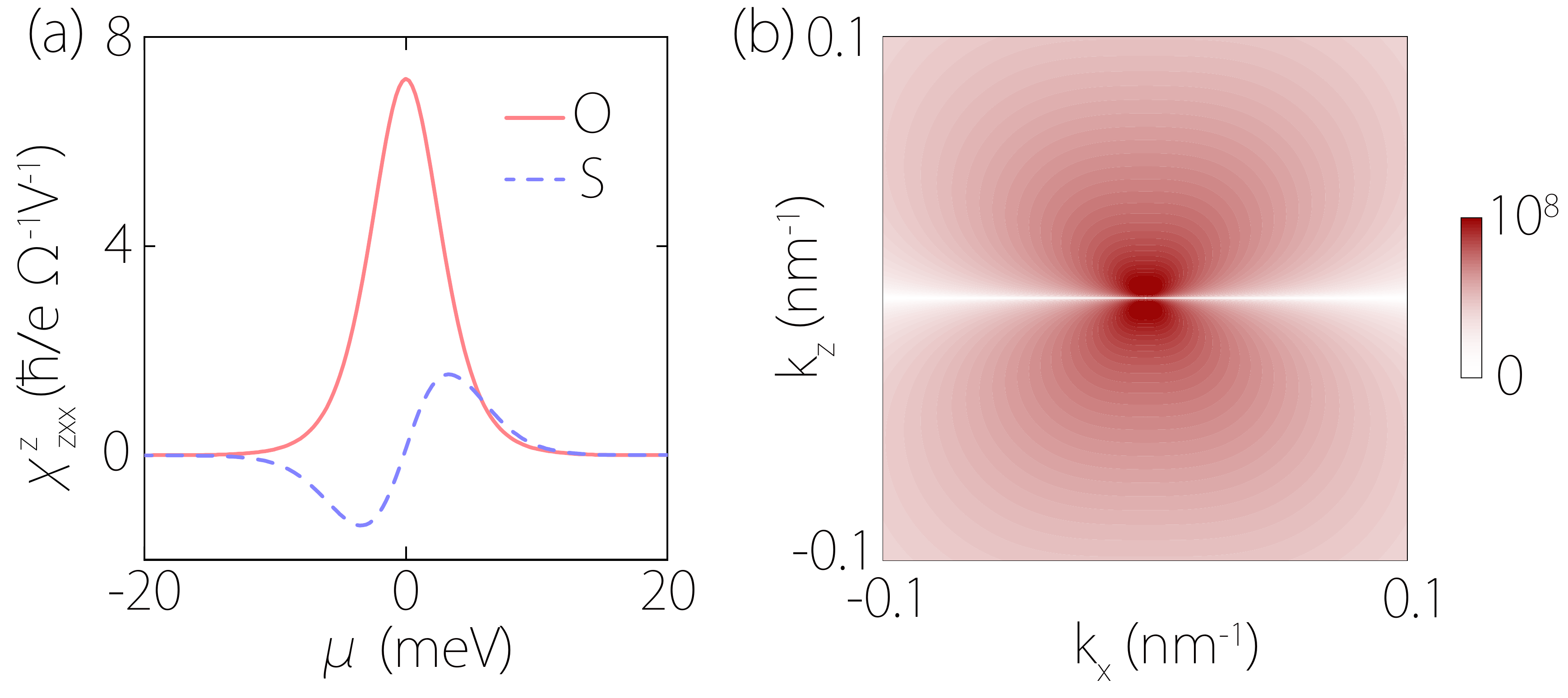} \caption{Nonlinear CP orbital current response in Weyl model. (a) Calculated intrinsic NOHE conductivity $\chi_{zxx}^z$ (red curve) versus chemical potential $\mu$. The blue dashed curve is for the corresponding nonlinear spin current response. (b) $k$-resolved contribution to $\chi_{zxx}^z$ for $\mu$ slightly above the Weyl point. The unit of color map is $\mathrm{nm^{3}/V}$. Here, we take $m=1.2 m_e$, $\nu=1.0\ \text{eV}\cdot \text{\AA}$, and $T = 20$ K.
}%
\label{fig_model}%
\end{figure}

To illustrate the general features, we first apply this formula to a Weyl point model:
\begin{equation}\label{mm}
  \mathcal{H}=\frac{k^2}{2m}+\nu \bm k\cdot\bm \sigma,
\end{equation}
where $m$ is an effective mass, $\bm \sigma$ is the vector of Pauli matrices, and $\nu=\pm 1$ determines the chirality of the Weyl point at $k=0$.
This model permits a CP nonlinear orbital current $j^z_z$ driving by applied $E$ field in the $x$-$y$ plane.
Figure~\ref{fig_model}(a) shows the corresponding tensor component $\chi^{z}_{zxx}$ calculated as a function of chemical potential $\mu$.
One can see the NOHE response is peaked at the Weyl point energy, showing a significant enhancement by topological band degeneracy. In Fig.~\ref{fig_model}(b), we plot the $k$-resolved contribution (the integrand of (\ref{key1})) to $\chi^{z}_{zxx}$ for $\mu$ slightly above the Weyl point,
confirming the enhancement is indeed from Weyl electron states. For comparison, by taking $\sigma$ in model (\ref{mm}) as spin, we also evaluate the nonlinear spin Hall conductivity, plotted as the dashed curve in Fig.~\ref{fig_model}(a).
One observes that (i) although the spin response is also enhanced around the Weyl point, its magnitude is still much smaller than the orbital one (note the Weyl model is a model with strong SOC); (ii) the spin response exhibits a (resonance-like) line-shape different from the peak profile of orbital response. These results are consistent with the general features we argued above.

\textcolor{blue}{\textit{RhSi: weak SOC.}}
Since NOHE does not require SOC, we first investigate it in a topological metal with weak SOC: RhSi, which has a chiral structure with $T$ point group symmetry. Our calculation reveals it can produce a remarkable out-of-plane torque generated by NOHE, stronger than the linear out-of-plane spin-orbit torques reported so far.

\begin{figure}[t!]
\centering
\includegraphics[width=0.48\textwidth]{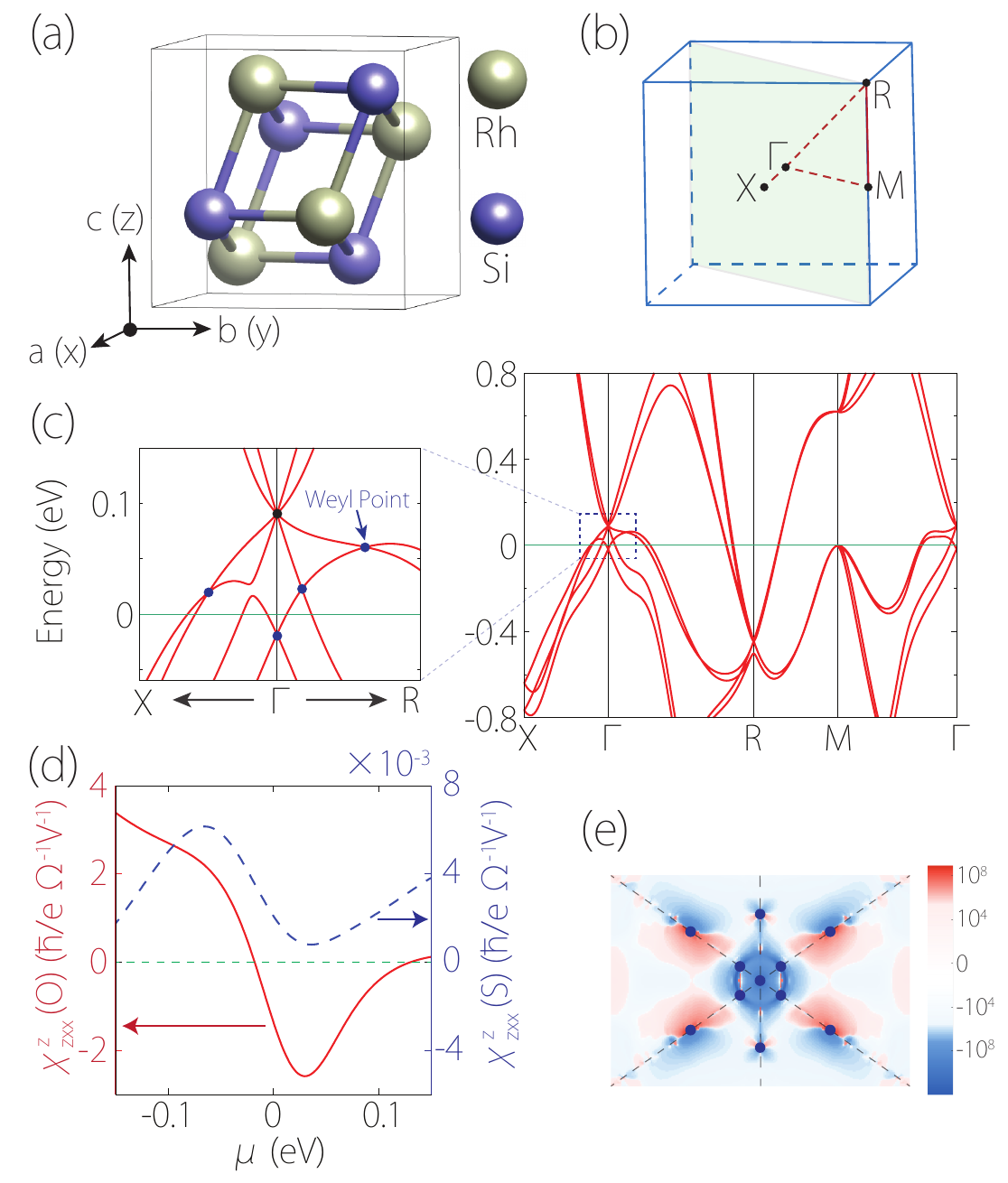} \caption{(a) Lattice structure and (b) Brillouin zone of $\mathrm{RhSi}$. (c) Calculated band structure with SOC. Left panel shows the enlarged view of the dashed box around $\Gamma$. (d) Calculated NOHE conductivity $\chi_{zxx}^z$(O) (red curve) versus chemical potential at room temperature. The blue dashed curve is for the corresponding nonlinear spin current response $\chi_{zxx}^z$(S).
 (e) $k$-resolved contributions to NOHE $\chi_{zxx}^z$ around the $\Gamma$ point on the intrinsic Fermi level in $\Gamma$-M-R plane. The unit of color map is $\mathrm{\AA^{3}/V}$.
}%
\label{fig_RhSi}%
\end{figure}

The lattice structure of RhSi is shown in Fig.~\ref{fig_RhSi}(a). The CP orbital Hall current is forbidden in linear response.
However, according to Table~\ref{materials_symmetry}, it is allowed by NOHE, characterized by components $\chi_{zxx}^{z}=\chi_{xyy}^{x}=\chi_{yzz}^{y}$ and $\chi_{zyy}^{z}=\chi_{yxx}^{y}=\chi_{xzz}^{x}$ (the coordinate axis are marked in Fig.~\ref{fig_RhSi}(a)). The calculated band structure of RhSi is plotted in Fig.~\ref{fig_RhSi}(c), which is consistent with previous works~\cite{chang2017,ni2020} (Calculation details are given in Supplementary Note 8). There are several band crossing points close to Fermi level. At the $\Gamma$ point, there are two crossing points with twofold and fourfold degeneracy. Along high-symmetry paths, such as $\Gamma$-X and $\Gamma$-R, RhSi also host type-I and type-II Weyl points, as shown in Fig.~\ref{fig_RhSi}(c).
As we mentioned, such topological metal band structure is promising to trigger large NOHE.

Figure~\ref{fig_RhSi}(d) shows the calculated $\chi_{zxx}^z$ of NOHE as a function of chemical potential at room temperature. For comparison, the corresponding component of nonlinear spin Hall response is also evaluated (dashed line).
One finds that the nonlinear orbital Hall conductivity is large $\sim -1.35$ $\hbar/e$ $\Omega^{-1}\mathrm{V}^{-1}$ at intrinsic Fermi level, and it is nearly three orders of magnitude larger than the spin response.
To pinpoint the origin of this large nonlinear orbital response, we plot the $k$-resolved contribution to $\chi_{zxx}^z$ in Fig.~\ref{fig_RhSi}(e), which shows significant contributions are coming from the regions around the band degeneracy points. These results demonstrate that the NOHE response is greatly enhanced by topological nodal features of RhSi.


To compare NOHE with previously reported linear spin and orbital Hall effects, it is convenient to convert $\chi$ into an effective linear out-of-plane orbital Hall conductivity, defined by $\sigma_{\mathrm{OH}}^{\mathrm{oop}} = \chi \rho j$,
where $j$ is the injected charge current density and $\rho$ is the (linear) longitudinal resistivity of the material.
Using the reported value $\rho \sim 3 \times 10^{-4}\ \Omega\ \mathrm{cm}$ for RhSi~\cite{Maulana2020} and $j = 10^7\ \mathrm{A/cm^2}$, we find $\sigma_{\mathrm{OH}}^{\mathrm{oop}}  \sim -4050\ \hbar/e\ \Omega^{-1}\mathrm{cm}^{-1}$.
Furthermore, in order for orbital current to produce a torque on a neighboring magnetic layer, the orbital moment has to be converted into spin moment after entering the magnetic layer, characterized by an orbital-to-spin conversion efficiency $\eta$. Thus, concerning the resulting torque, it is more meaningful to consider the combination $\eta \sigma_{\mathrm{OH}}^{\mathrm{oop}}$.
For $3d$ ferromagnets, such as Fe, Co, CoFe, and Ni, $\eta$ is positive and ranges from 0.92\% to 4.55\%~\cite{lee2021orbital}. More recent works found larger $\eta$ in Gd ($\eta \sim -18\%$)~\cite{DingPRL2024} and Fe$_3$GaTe$_2$ ($\eta\sim 38\%$) \cite{Jiang2025orbital-FGT}. For orbital torques exerted on these two materials, the value of $\eta \sigma_{\mathrm{OH}}^{\mathrm{oop}}$ can reach $729\ \hbar/e\ \Omega^{-1}\mathrm{cm}^{-1}$ and $-1539\ \hbar/e\ \Omega^{-1}\mathrm{cm}^{-1}$, respectively. Such values are much larger than the out-of-plane torques (using low-symmetry materials) reported to date, which are usually much less than $200\ \hbar/e\ \Omega^{-1}\mathrm{cm}^{-1}$~\cite{Ralph2017WTe2,Yang2023TaIrTe4,Yu2023TaIrTe4,Wang2023MnPd3,Dash2024TaIrTe4,Yang2024,Sheng2024}.

In addition, the torque efficiency is typically characterized by the effective orbital/spin Hall angle. It is defined as $\theta_{\mathrm{OH}}^{\mathrm{oop}}=\eta(2e/\hbar)\rho\sigma_{\mathrm{OH}}^{\mathrm{oop}}$ for NOHE, and it can reach 0.44 and -0.93 for RhSi to produce orbital torques on Gd and Fe$_3$GaTe$_2$, respectively. These values are significantly larger than the ever reported strongest out-of-plane linear torque efficiencies, which range from 0.03 to 0.11~\cite{Yang2023TaIrTe4,Yu2023TaIrTe4,Wang2023MnPd3,Dash2024TaIrTe4,Yang2024,Loh2024}.

{\renewcommand{\arraystretch}{1.2}
\begin{table}[pt]
\caption{{Results of nonlinear CP orbital current and spin current responses for the representative topological semimetals.} Calculated nonlinear orbital Hall conductivity (NOHC) and nonlinear spin Hall conductivity (NSHC) are in units of $\hbar/e$ $\Omega^{-1}\mathrm{V}^{-1}$. OHA and SHA denote the orbital Hall angle and spin Hall angle, respectively. And the last three lines are for the out-of-plane effective orbital Hall angle including the $\eta$ factor, for $\eta=2.91\%$ (Co), $\eta=-18\%$ (Gd) and $\eta=38\%$ (Fe$_3$GaTe$_2$). }%
\label{comparison}%
\setlength{\tabcolsep}{2.2mm}{
\begin{tabular}
[b]{ccccc}\hline\hline
 & \ \ RhSi\ \ &
\ \  YPtBi \ \ & \ \ $\mathrm{PbTaSe_{2}}$\ \
\\\hline
NOHC\ \  & -1.35 & \ \ -1.58 & -0.76 & \\
\ NSHC\ \  & 0.0021 & \ \ 0.020 & 0.023 & \\
OHA\ \ & -2.44 & \ \ -2.84 & -3.63 & \\
SHA\ \  & 0.0038 & \ \ 0.036 & 0.11 & \\
$\theta_{\mathrm{OH}}^{\mathrm{oop}}$ $(\eta=2.91\%)$\ \  & -0.071 & \ \ -0.083 & -0.11 & \\
$\theta_{\mathrm{OH}}^{\mathrm{oop}}$ $(\eta=-18\%)$\ \  & 0.44 & \ \ 0.51 & 0.65 & \\
$\theta_{\mathrm{OH}}^{\mathrm{oop}}$ $(\eta=38\%)$\ \  & -0.93 & \ \ -1.08 & -1.38 & \\
\hline\hline
\end{tabular}}
\end{table}}



\textcolor{blue}{\textit{YPtBi: strong SOC and switchable orbital torque.}}
Next, we consider a topological metal with strong SOC: YPtBi. We will see that NOHE still dominates over its spin counterpart, due to the enhancement from band degeneracy.

\begin{figure}[t]
\centering
\includegraphics[width=0.48\textwidth]{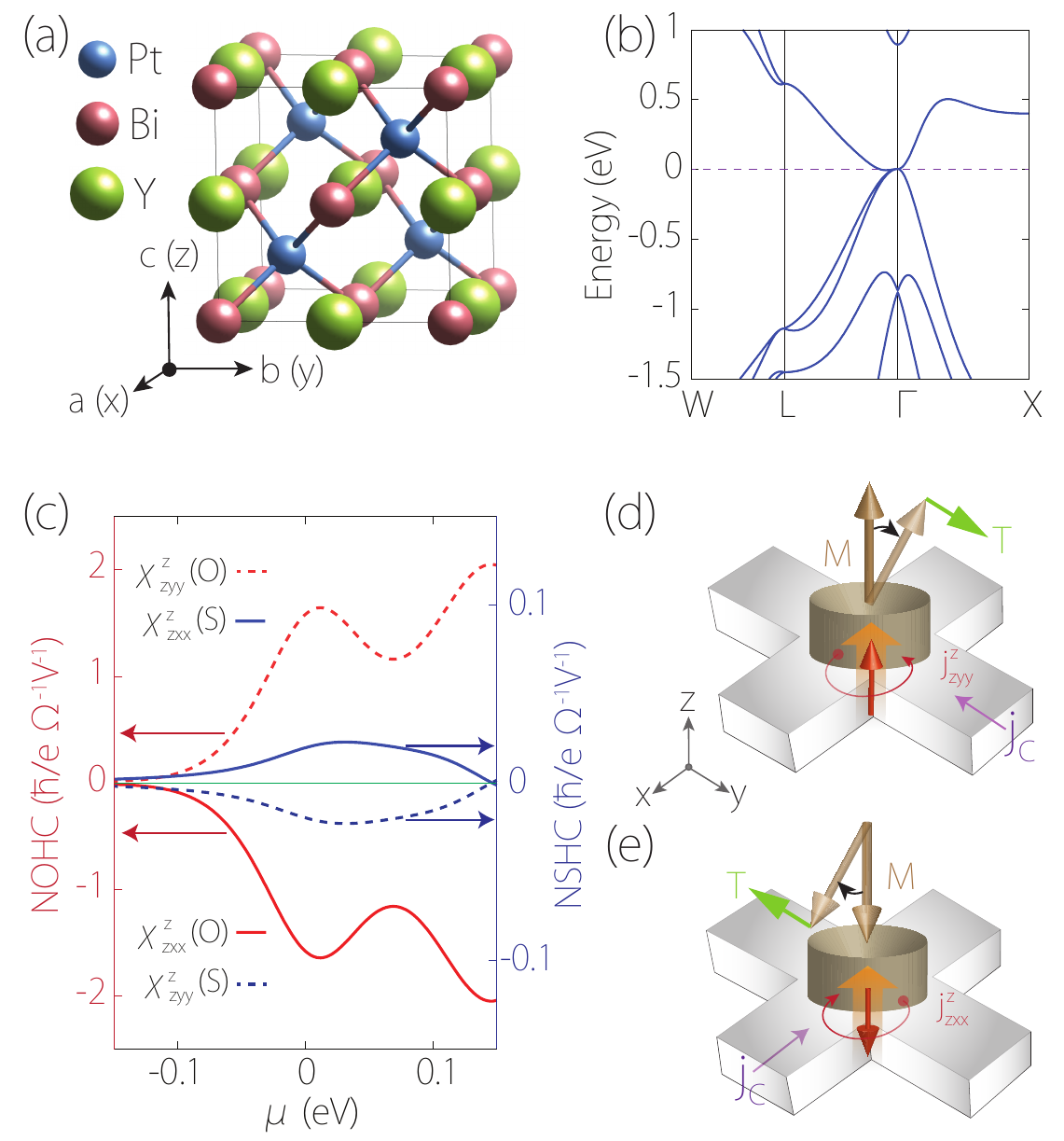}
\caption{Switchable CP NOHE in YPtBi. (a) Lattice structure of $\mathrm{YPtBi}$. (b) Band structure with SOC included. (c) Calculated NOHE conductivities $\chi_{zxx}^z$ and $\chi_{zyy}^z$ (red curves) versus chemical potential at room temperature.
The  blue curves show the results for the corresponding nonlinear spin current responses. 
(d,e) Schematic figure for a bilayer device structure. 
The CP orbital current generated from NOHE in the bottom layer (and the resulting torque on the top magnetic layer) can be switched by controlling the driving current direction (in $x$ or $y$ direction).}
\label{fig_YPtBi}%
\end{figure}

YPtBi has attracted wide interest for potential applications in superconductivity, thermoelectricity, and spintronics~\cite{Butch2011, Goyal2021,Shirokura2024}. As shown in Fig.~\ref{fig_YPtBi}(a), it is a half-Heusler crystal with space group $F\bar{4}3m$  (No.~216). The calculated band structure of YPtBi is plotted in Fig.~\ref{fig_YPtBi}(b). It is semimetallic, with a fourfold Dirac point near the Fermi level at $\Gamma$.
The result is consistent with previous studies~\cite{feng2010}.

The $T_d$ point group symmetry of YPtBi forbids any CP orbital Hall current in linear response, but allows it at nonlinear order. To generate CP orbital current flowing along the $z$ direction, the relevant
conductivity components are $\chi_{zxx}^z = -\chi_{zyy}^z$.
Figure~\ref{fig_YPtBi}(c) presents the calculated $\chi_{zxx}^z$ and $\chi_{zyy}^z$. The corresponding nonlinear spin Hall
conductivity components are also shown for comparison. One can see $\chi_{zxx}^z$ ($= -\chi_{zyy}^z$) exhibits a peak at intrinsic Fermi level, with a peak value of $-1.58\ \hbar/e\ \Omega^{-1}\mathrm{V}^{-1}$, manifesting the enhancement by degeneracy between conduction and valence bands.
And the orbital response is still two order of magnitude larger than the spin response ($\sim 0.020\ \hbar/e\ \Omega^{-1}\mathrm{V}^{-1}$), despite the spin response is already amplified by the large SOC.  These findings confirm the general features we argued above.

Using the experimentally measured resistivity $\rho \sim 3 \times 10^{-4}\ \Omega\ \mathrm{cm}$ at room temperature~\cite{Butch2011} and an injected current density $j = 10^7\ \mathrm{A/cm^2}$, we estimate the corresponding effective linear out-of-plane orbital Hall conductivity $\sigma_{\mathrm{OH}}^\mathrm{oop} \sim -4740\ \hbar/e\ \Omega^{-1}\mathrm{cm}^{-1}$, which is even larger than that of RhSi.


For applications, it is desirable to have an easy control over the sign of generated torque.
For NOHE, this cannot be done by simply reversing the sign of applied $E$ field (or injected current). However, for YPtBi,
because $\chi_{zxx}^z = -\chi_{zyy}^z$, the sign switching can be readily achieved by rotating the injected current direction by $\pi/2$. As illustrated in Figs.~\ref{fig_YPtBi}(d) and~\ref{fig_YPtBi}(e), the sign of the generated CP orbital current (and hence the orbital torque
on the top ferromagnetic layer) is opposite between the two in-plane current directions. This offers a simple
device setup to realize switchable orbital torques.

\textcolor{blue}{\textit{Discussion.}}
We have proposed a strategy to achieve field-free deterministic switching of PMA magnets, using the orbital torques generated by NOHE. As a new physical effect, we find NOHE permits the generation of CP orbital current, a key ingredient for
switching PMA magnets, in a large class of crystals, much beyond the scope allowed by linear response.
Moreover, we have shown that NOHE is significantly enhanced by band near-degeneracies, around which it can dominate over the spin response by orders of magnitude, even in systems with strong SOC. So far, studies on orbital torques mostly focused on materials with weak SOC, such as the $3d$ transition metals. By harnessing the nonlinear mechanisms, our findings greatly broadens the material platforms for orbitronics.

Particularly, our results suggest topological metals as promising orbital source materials, due to the dramatic enhancement of nonlinear orbital response by the interband coherence associated with topological band degeneracies.
Besides RhSi and YPtBi having nodal points in band structures, the similar feature is also manifested in nodal-line topological metals, such as PbTaSe$_2$ (see Table, details in Supplementary Note 5).


In the calculations, we focused on the intrinsic NOHE. At the level of relaxation time approximation, we obtain an extrinsic contribution to NOHE given by
\begin{equation}
  \chi^p_{abc}=-\tau^2\int [d\bm k]\partial_b f_0 \ \partial_c\langle j^p_a\rangle_n,
\end{equation}
where $\tau$ is the relaxation time. We have computed this contribution and found it is subdominant compared to the intrinsic NOHE for the material example RhSi (see Supplementary Note 7).
We note that although a systematic theory for the extrinsic NOHE remains to be developed, the general features, including the symmetry conditions, the enhancement by band near-degeneracies, and the comparison between orbital and spin responses, should hold for both intrinsic and extrinsic contributions.

\bibliography{ref}



\bigskip
\onecolumngrid
\begin{center}\textbf{End Matter}\end{center}
\twocolumngrid
\vspace{0.5\baselineskip}

\appendix
\renewcommand{\theequation}{A\arabic{equation}} 
\setcounter{equation}{0} 
\renewcommand{\thefigure}{A\arabic{figure}}
\setcounter{figure}{0}

{\emph{Appendix: Formulae of intrinsic NOHE.}}
The formulae for intrinsic NOHE can be obtained by the approach of extended semiclassical theory \cite{gao2014,gaoPRB2015,gao2019,xiao2021adiabatic}. In this approach, the orbital current density is expressed as
\begin{equation}\label{sc}
  j_a^p=\int [d\bm k] f_{n\bm k} \langle W_{n\bm k}|\hat{j}^p_a |W_{n\bm k}\rangle,
\end{equation}
where $[d\bm k]$ is a shorthand notation for $\sum_n d\bm k/(2\pi)^3$ with $n$ as the band index and $\bm k$ the crystal momentum, the last factor is the orbital current carried by the wave packet $|W_{n\bm k}\rangle$ which in the absence of $E$ field is centered at Bloch state $|u_{n\bm k}\rangle$, and $f_{n\bm k}$ is the distribution function.
For an intrinsic response, we take Fermi-Dirac distribution for $f_{n\bm k}$, and
the last factor in (\ref{sc}) can be obtained from a variation of semiclassical action, as developed in Refs. \cite{dong2020,xiao2021adiabatic} and detailed in
the Supplementary Note 3.

The intrinsic nonlinear orbital conductivity is obtained to be
%
%
%
\begin{equation}\begin{split}\label{key}
  \chi_{a(bc)}^{p}=&\frac{1}{2}\int [d\bm k] \Big[\Lambda_{abc,n}^{p}f_0
  \\
  &-\Big(\langle j_a^p\rangle_n G_{bc,n}-\langle v_b\rangle_n \mathfrak{G}_{ac,n}^p-\langle v_c \rangle_n \mathfrak{G}_{ab,n}^p\Big)f_0'\Big].
  \end{split}
\end{equation}
Here, we suppressed the $\bm k$ labels in the integrand, $f_0\equiv f_0(\varepsilon_n)$ is the Fermi distribution with
$\varepsilon_n$ the unperturbed band energy; $\langle v_a\rangle_n=\langle u_{n}|\hat{v}_a|u_{n}\rangle$ is the intraband velocity matrix element, similar for $\langle j_a^p\rangle_n$.
%
Besides,
\begin{equation}\label{k-BCP}
  G_{bc,n}=2 \text{Re}\sum_{\ell \neq n}\frac{\langle v_b\rangle_{n\ell}\langle v_c\rangle_{\ell n}}{(\varepsilon_{n}-\varepsilon_{\ell})^3}
\end{equation}
is known as the Berry connection polarizability (BCP) \cite{gao2014,liu2022third}, which already plays important roles in several nonlinear effects \cite{wang2021,liu2021,lai2021,Gao2023QM}, and
\begin{equation}\label{m-BCP}
  \mathfrak{G}_{ac,n}^p=2 \text{Re}\sum_{\ell\neq n}\frac{\langle j_{a}^{p}\rangle_{n\ell}\langle v_c\rangle_{\ell n}}{(\varepsilon_{n}-\varepsilon_{\ell})^3}
\end{equation}
may be dubbed the orbital BCP.
In the same sense,
\begin{widetext}
\begin{align}
\Lambda_{abc,n}^{p} =-2\mathrm{{\operatorname{Re}}}\sum_{m\neq n}&\bigg[\frac
{3\langle v_{b}\rangle_{nm}\langle v_{c}\rangle_{mn}\left(\langle j_{a}^{p}\rangle_n-\langle j_{a}^{p}\rangle_m \right)
}{(\varepsilon_{n}-\varepsilon_{m})^{4}}
-\frac{\langle \partial_{b} {j}_{a}%
^{p}\rangle_{nm}\langle v_{c}\rangle_{mn}+\langle \partial_{c}{j}_{a}%
^{p}\rangle_{nm}\langle v_{b}\rangle_{mn}}{(
\varepsilon_{n}-\varepsilon_{m})  ^{3}} \nonumber\\
& -\sum_{\ell\neq n} \frac{\left(
\langle v_{b}\rangle_{\ell m}\langle v_{c}\rangle_{mn}+\langle v_{c}\rangle_{\ell m}\langle v_{b}\rangle_{mn}\right) \langle j_{a}^{p}\rangle_{n\ell}}{(  \varepsilon_{n}-\varepsilon_{\ell})  (\varepsilon
_{n}-\varepsilon_{m})^{3}}
-\sum_{\ell\neq m} \frac{\left(
\langle v_{b}\rangle_{\ell n}\langle v_{c}\rangle_{nm}+\langle v_{c}\rangle_{\ell n}\langle v_{b}\rangle_{nm}\right) \langle j%
_{a}^{p}\rangle_{m\ell}}{(  \varepsilon_{m}-\varepsilon_{\ell})  (\varepsilon
_{n}-\varepsilon_{m})^{3}}\bigg]
\end{align}
\end{widetext}
can be regarded as the orbital counterpart of BCP dipole $\partial_a G_{bc}$, where $\partial_{a                         }\equiv \partial/\partial{{k}_{a}}$.



\end{document}